# Anisotropic magnetization of RuEu$_{1.5}$Ce$_{0.5}$Sr$_2$Cu$_2$O$_{10}$ (Ru-1222) thin films

I. Felner[a] G. Kopnov[a] and G. Koren[b]

.

[a] The Racah Institute of Physics, The Hebrew University of Jerusalem, Jerusalem 91904, Israel
[b] Department of Physics, Technion – Israel Institute of Technology, Haifa 32000, Israel

A dc magnetic study is reported on c-axis oriented epitaxial thin films of EuSr$_2$Eu$_{1.5}$Ce$_{0.5}$Cu$_2$O$_{10}$ (Ru-1222) on (100) SrTiO$_3$ wafers. Magnetic measurements on 200 and 500 nm thick films were performed with the magnetic field either parallel or perpendicular to the wafer. We found that the films order magnetically at 180 K. The easy axis of the magnetization is in the basal plane and the Ru$^{5+}$ ions are in their high-spin state, as deduced from the isothermal magnetization curves at low temperatures. The anisotropy ratio was found to be 7.5. The observed results are compared to data of the Ru-1222 ceramic material.

## 1. INTRODUCTION

Much current attention is given to the tetragonal rutheno-cuprate layered RuSr$_2$R$_{2-x}$Ce$_x$Cu$_2$O$_{10}$ (R=Eu and Gd, Ru-1222) compounds in which coexistence of magnetic order at T$_M$= 125-170 K and superconductivity (SC) below T$_C$ = 32-50 K was discovered [1]. Both T$_M$ and T$_C$ (T$_M$ >T$_c$) depend on oxygen concentration and sample preparation. The SC charge carriers originate from the CuO$_2$ planes and the hole doping of these planes, can be optimized with appropriate variation of the R/Ce ratio [2]. The magnetic state which is confined to the Ru layers remains unchanged and persists when SC sets in. X-ray-absorption near- edge spectroscopy (XANES) measurements, taken at the K edge of Ru, reveal that the Ru ions are basically pentavalent irrespective of the Ce concentration [3].

The SC state in the Ru-1222 system is well established and understood by now, while determination of the magnetic structure of the Ru-O layers has not been published so far. Generally speaking, the M/H(T) dc magnetic studies exhibit two magnetic transitions at $T_{2M}$ (around 80-90 K) and at $T_M$ ($T_{2M} < T_M$). At low applied magnetic fields (H<2-3 kOe), irreversibility in the zero-field-cooled (ZFC) and field-cooled (FC) curves is observed up to $T_{2M}$ when these two curves merge. For higher external fields (H>5 kOe), both ZFC and FC curves collapse to a single ferromagnetic-like behavior. The M/H(T) curves do not enable an easy determination of $T_M$ which is generally obtained directly from the temperature dependence of the saturation moment ($M_{sat}$), and also from Mossbauer studies (MS) on $^{57}$Fe doped materials.[1] In the M(H) curves, a relatively wide ferromagnetic (FM) hysteresis loops are opened up at low temperatures with a coercive field, $H_C$ of ~450-500 Oe at 5 K. These loops become narrower as the temperature increases and practically close down below $T_{2M}$. Above $T_{2M}$, small canted AFM-like hysteresis loops are observed and the corresponding $H_C(T)$ curves show a peak with a maximum at around 120 K ($H_C$ ~150 Oe) and become zero at $T_M$ [4].

Since single crystals of Ru-1222 are unavailable at the present time, all magnetic investigations on this compound were so far limited to ceramic polycrystalline materials. Hence, the experimental magnetic results represent average values only. The recent first report on highly c-oriented Ru-1222 thin films by W.Q. Lu et. al.[5] exhibits coexistence of SC at $T_C$=32 K, and magnetic ordering around 130 K. However, a detailed study of magnetic properties including determination of the magnetic easy axis in the films, (which may shed light on the magnetic structure) is absent.

In the present study we measured the magnetic behavior of highly c-axis oriented epitaxial RuEu$_{1.5}$Ce$_{0.5}$Sr$_2$Cu$_2$O$_{10}$ thin films in both parallel (H$_{PAR}$) and perpendicular (H$_{PERP}$) fields to the film surface. We find that the easy axis is located along the basal plane, and that M$_{sat}$ is ~3 $\mu_B$, (at 5 K). This value agrees well with the theoretical moment of Ru$^{5+}$ in its the high spin state. $M_{sat}$ parallel to the c-direction is found to be ~0.4 $\mu_B$, indicating that the anisotropy factor for Ru-1222 is ~7.5.

## 2. EXPERIMENTAL DETAILS

The ceramic target from which the films were prepared had a nominal composition of $RuSr_2Eu_{1.5}Ce_{0.5}Cu_2O_{10}$. It was prepared by a solid-state reaction technique under oxygen atmosphere as follows. Prescribed amounts of $Eu_2O_3$, $CeO_2$, $SrCO_3$, CuO and Ru were mixed, pressed into pellets and preheated at 950° C for 1 day. The product was cooled, reground and sintered at 1050° C for 2 days, and then furnace cooled. The films were deposited on (100) $SrTiO_3$ (STO) wafers by 10ns long laser pulses of 355nm wavelength produced by the third harmonic of a Nd-YAG laser. The properties of two different films are reported in the present study. The first film of 200nm thickness, was deposited by 2000 laser pulses of 1.5 J/cm$^2$ laser fluence on the target at a wafer temperature of 790° C and under 100 mTorr of flowing oxygen. Annealing was done by slow cooling to room temperature at a rate of 50° C/h in 0.8 atm of $O_2$. The second 500 nm thick film, was deposited by 5000 pulses of 4 J/cm$^2$ fluence, at 745° C wafer temperature and under 400 mTorr $O_2$ flow. This film was annealed at 510° C under 0.8 atm $O_2$, with a dwell time of 1 h. None of the films was found to be superconducting.

X-ray diffraction (XRD) analysis was performed to verify the films purity and orientation. Within the instrumental accuracy, the two films studied have a tetragonal structure (space group I4/mmm) with the same a=3.835(5)Å lattice parameter. Isothermal magnetization, ZFC and FC dc magnetic measurements in the range of 5-250K were performed in a commercial (Quantum Design) super-conducting quantum interference device (SQUID) magnetometer. The microstructure and the phase integrity of the materials were investigated by QUANTA (Fri Company) scanning electron microscopy (SEM) and by a Genesis energy dispersive x-ray analysis (EDAX) device attached to the SEM.

## 2. EXPERIMENTAL RESULTS

We found that our Ru-1222 films deposited on STO substrate grow with c-axis orientation normal to the wafer, as shown in Fig. 1. The 200 and 500-thick films' area are 0.36(1) and 0.27(1)cm$^2$, respectively. The strong peak other than the marked Ru-1222 (00l) reflections is assigned to the STO substrate. The absence of any other non (00l) reflections indicates clearly that the film is perfectly c-axis oriented. The c lattice parameter deduced from the XRD patterns are 28.36(2) and 28.50(2)Å for the 200 and 500 nm-thick films, respectively in good agreement with the value obtained for sintered ceramic materials[1]. The morphology detected by the SEM, shows a density packed structure with a rough and uniform surface. EDAX analysis is close to the initial stoichiometric composition. AFM measurements revealed good crystallinity with uniform directionality (Fig. 2) over areas of a few thousands of µm$^2$ and therefore the films are not only c-axis oriented but also fully apitaxial. None of the studied films is superconducting even after long annealing under oxygen at elevated temperatures. Post annealing of the films at 1050° C under oxygen and slow cooling to ambient temperature did not induce SC. Since in Ru-1222 the SC state depends strongly on the oxygen concentration, we may assume that the full epitaxiallity and the perfect crystallinity of the present films hinder the oxygen uptake, thus the films are in the under-doped region. Since the main aim of the present paper is the determination of the easy magnetic axis, the present well oriented c-axis films permit such a study. For the sake of brevity, we'll present only magnetic data obtained for the 200 nm-thick film.

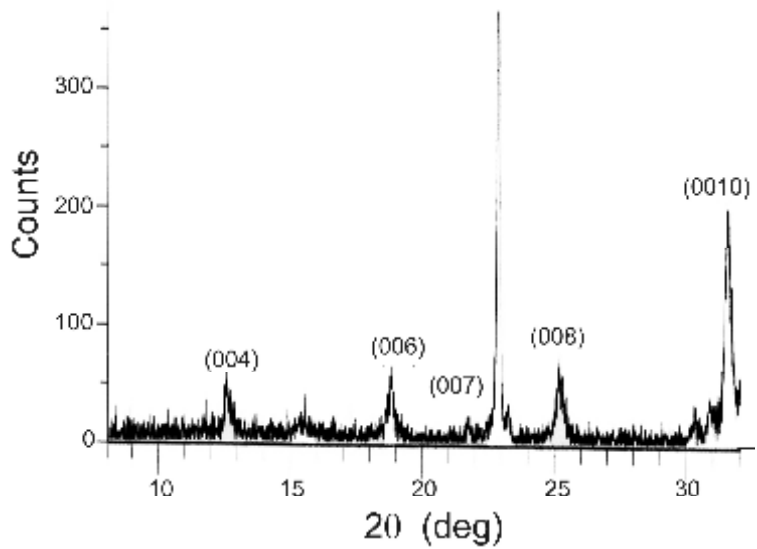

Figure 1. XRD pattern for the 500 nm thick RuEu$_{1.5}$Ce$_{0.5}$Sr$_2$Cu$_2$O$_{10}$ thin film. The unmarked reflection belongs to the STO substrate

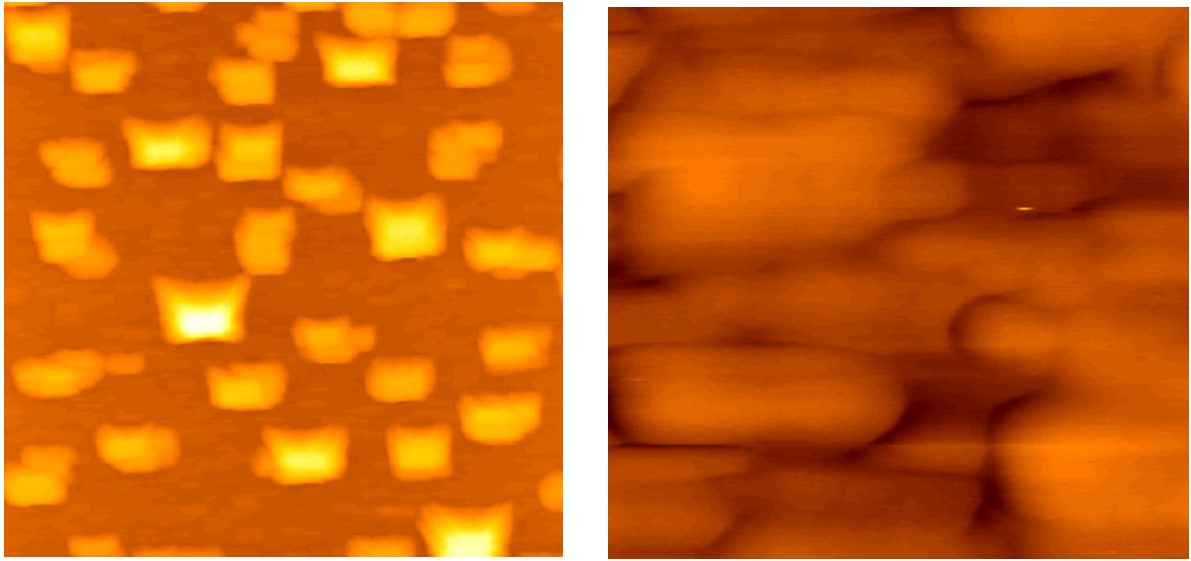

Figure 2. AFM topographic images of the 200 nm-thick $RuEu_{1.5}Ce_{0.5}Sr_2Cu_2O_{10}$ thin film. The areas are 5x5 $\mu m^2$ (left panel) and  0.25x0.25 $\mu m^2$ (right panel)

The ZFC and FC curves measured at $H_{PAR}$ and $H_{PERP}$ of 90 Oe, are presented in Fig. 3. One definitely sees the irreversibility of the two branches which merge in both orientations at $T_M$ around 180 K. This $T_M$ value is somewhat higher than that obtained for non-superconducting ceramic Ru-1222 bulk material [2].

Isothermal M(H) measurements at various temperatures in both directions have been carried out, and the results obtained for 5 K are exhibited in Fig. 4. All M(H) curves  below  $T_M$, are strongly dependent on the field (up to  2-3 kOe). At higher fields the magnetization can be described as: M(H)= $M_{sat}$+ $\chi$H- $\chi^*$H, where $M_{sat}$ corresponds to the magnetic contribution of the Ru sub lattice, $\chi$H is the small linear paramagnetic moment of Eu and Cu (which is neglected here) and -$\chi^*$H is the large diamagnetic contributions of  STO substrate. The saturation moments obtained at 5 K are $M_{sat} = 3.0(2)$ and $0.40(3)$ $\mu_B$ for $H_{PAR}$ and $H_{PERP}$, respectively. Due to the small magnitude of the magnetic moments, no corrections for demagnetization contributions were made. However, since the saturation moments are independent of field, the demagnetization factor does not affect the values of these moments. Similar M(H) curves have been measured at various temperatures. However, when the temperature increases, the Ru contribution decreases, whereas the diamagnetic signal remains practically unchanged. Therefore any determination of $M_{sat}$ at higher temperatures is misleading. Three significant conclusions can be made from the results in Fig. 3. (i) The higher magnetic moment value obtained for $H_{PAR}$ indicates that the easy axis of the magnetization in Ru-1222 is confined to the basal plane. This is consistent with our previous easy axis determination based on Mossbauer spectroscopy measurements of dilute [57]Fe doped Ru-1222 ceramic materials.[1] (ii) The measured $M_{sat} = 3$ $\mu_B$ value, is identical to the fully saturated $3\mu_B$ moment expected for $Ru^{5+}$ in the high-spin state, i.e. $g\mu_B S$ for $g=2$ and $S=1.5$. This also confirms the XANES results discussed above [5]. This value is higher than the average paramagnetic effective moment ($P_{eff} = 2.15$ $\mu_B$) obtained for Ru-1222 ceramic materials. (iii) The anisotropy factor of Ru-1222 for $H_{PAR}$ and $H_{PERP}$ is 7.5.

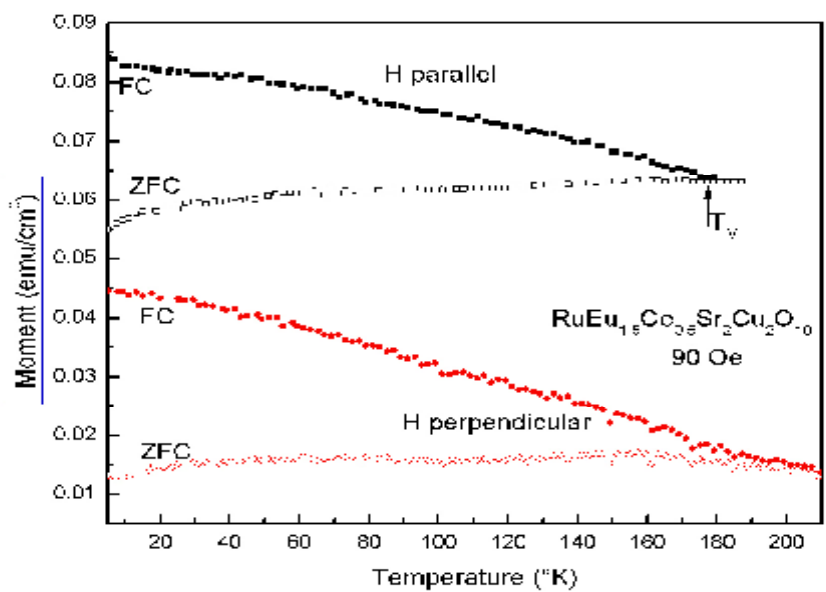

Figure 3. ZFC and FC moments versus temperature for the 200 nm-thick $RuEu_{1.5}Ce_{0.5}Sr_2Cu_2O_{10}$ thin film.

At low applied fields, the M(H) curve exhibits a typical ferromagnetic-like hysteresis loop (inset of Fig. 4) for both $H_{PAR}$ and $H_{PERP}$. The coercive field ($H_C = 500(40)$ Oe at 5 K) is similar to that reported for ceramic materials [1,2].

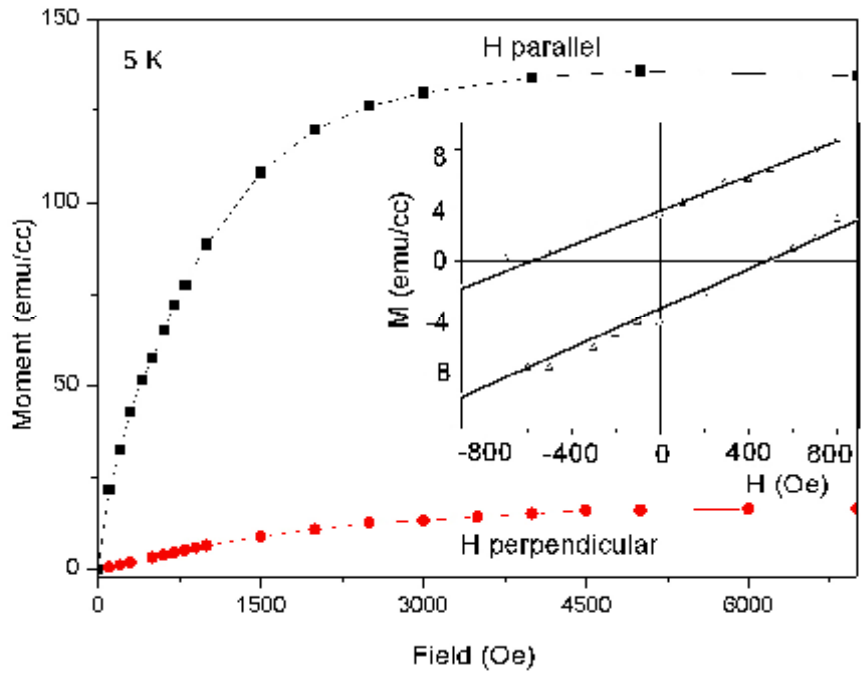

Figure 4. Magnetization curves at 5 K for the 200 nm-thick RuEu$_{1.5}$Ce$_{0.5}$Sr$_2$Cu$_2$O$_{10}$ film measured in both directions. The inset shows the low field hysteresis loop obtained at 5 K for H parallel.

## 4. CONCLUSIONS

highly oriented and epitaxial c-axis thins film of RuEu$_{1.5}$Ce$_{0.5}$Sr$_2$Cu$_2$O$_{10}$ on (100) SrTiO$_3$ wafers and measure their magnetic properties at low temperatures. The films exhibit the typical magnetic features assigned to the Ru-1222 system with $T_M = 180$ K. The field dependence of the magnetization in the low temperature limit indicates that the Ru$^{5+}$ ions in this compound are in their high-spin state, and that the easy axis of the magnetization is in the basal plane. Our results also show that the anisotropy ratio in the c-axis oriented Ru-1222 films is 7.5. Therefore, magnetism and superconductivity are mostly decoupled and can coexist in this compound, as the internal magnetic field is basically confined to the Ru-O layers, while superconductivity resides mostly in the Cu-O planes. The exact nature and local structure of the Ru layers is presently unknown, but neutron diffraction measurements could determine this structure precisely.

**Acknowledgments:** This research is supported in part by the Israel Science Foundation (ISF, 2004 grants numbers: 618/04 and 1565/04), and by the Klachky Foundation for Superconductivity.

**REFERENCES**

[1] I Felner, U. Asaf, Y. Levi, and O. Millo, Phys. Rev. B 55 (1997) R3374.

[2] I Felner, I. Asaf, U and E. Galstyan, Phys. Rev. B 66 (2002) 024503.

[3] G.V.M. Williams, L.-Y. Jang and R.S.Liu, Phys. Rev. B 65 (2002) 064508.

[4] I. Felner, V.P.S. Awana and E. Takayama-Muromachi, Phys. Rev. B 68 (2003) 094508.

[5] W.Q. Lu, Y. Yamamoto, I. Ohkubo, V.V. Petrykin, M. Kakihana, Y. Matsumoto and H. Koinuma, Physica C 405 (2004) 21.